\documentclass[seceq]{ptptex}

\usepackage{graphicx}

\usepackage{wrapft}

\pubinfo{Vol.~120, No.~5, November 2008}

\title{
Phenomenology of high gluon density QCD and heavy-ion physics at ISMD~2011:
 $x$ smaller than ever~!
}

\author{
Adrian \textsc{Dumitru}$^{1,2}$\footnote{e-mail: Adrian.Dumitru@baruch.cuny.edu}
}

\inst{
$^1$RIKEN BNL Research Center, Brookhaven National
  Laboratory\\ Upton, NY 11973, USA\\
$^2$Department of Natural Sciences, Baruch College, CUNY\\
17 Lexington Avenue, New York, NY 10010, USA
}

\recdate{
April 1, 2004; revised November 20, 2008}

\abst{
I provide a brief summary of the theory presentations at ISMD 2011
related to the phenomenology of small-$x$ QCD evolution and its
application to particle production and fluctuations in high-energy
hadron and heavy-ion collisions. I also mention some challenges for
quantitative phenomenology which emerged from the LHC, such as
understanding the long-range ``ridge'' in high-multiplicity $p+p$
collisions, the transverse momentum distributions in $p+p$ at
semi-hard $p_\perp$, and the origin and scale of density fluctuations
in the initial state of heavy-ion collisions.}

\begin{document}
\maketitle

\section{Introduction}
The physics of multi-particle dynamics has a long history. Indeed,
ISMD 2011 on Miyajima-Island, Hiroshima, was the 41$^\mathrm{st}$
meeting in the {\it International Symposium on Multiparticle Dynamics}
series. The talks have mostly focused on physics at accelerator
facilities, in particular at the BNL-RHIC and CERN-LHC colliders. They
covered a very broad range of topics including particle production and
multi-particle correlations in high-energy collisions, studies of hard
QCD jets and of the underlying event, collective effects in QCD at
high temperatures and baryon densities, femtoscopic Bose-Einstein
correlations, as well as electroweak physics of the standard model and
beyond. However, we had talks also on cosmic ray physics and its
relation to accelerator based high-energy QCD, and finally on
strong-field QED with high intensity lasers. It is, indeed, a time of
exciting progress in this field and ISMD~2011 provided a nice overview
as well as a platform for stimulating discussions.

This write-up is intended to offer a brief summary of the theory
presentations at ISMD 2011 related to small-$x$ QCD; more
specifically, of the phenomenology of small-$x$ QCD evolution and its
application to particle production and fluctuations
in high-energy hadron and heavy-ion collisions.

\section{Phenomenology of running-coupling BK evolution}

Collisions of protons or heavy ions such as lead at high energies
release a very large number of gluons from their wave functions which
dominate the production of new particles and anti-particles. In fact,
the wave function of a hadron boosted to (nearly) the light cone is so
densely packed with gluons that they may ``overlap'', leading to
non-linear interactions~\cite{GLR}. Therefore, at high energies the colliding
hadrons can be treated as a high occupancy gluon field. This dense
system is nowadays referred to as a Color Glass Condensate (CGC). For
a recent in-depth review of the CGC we refer to ref.~\cite{CGCrev}.

\begin{figure}
  \centerline{
\includegraphics[width=8cm]{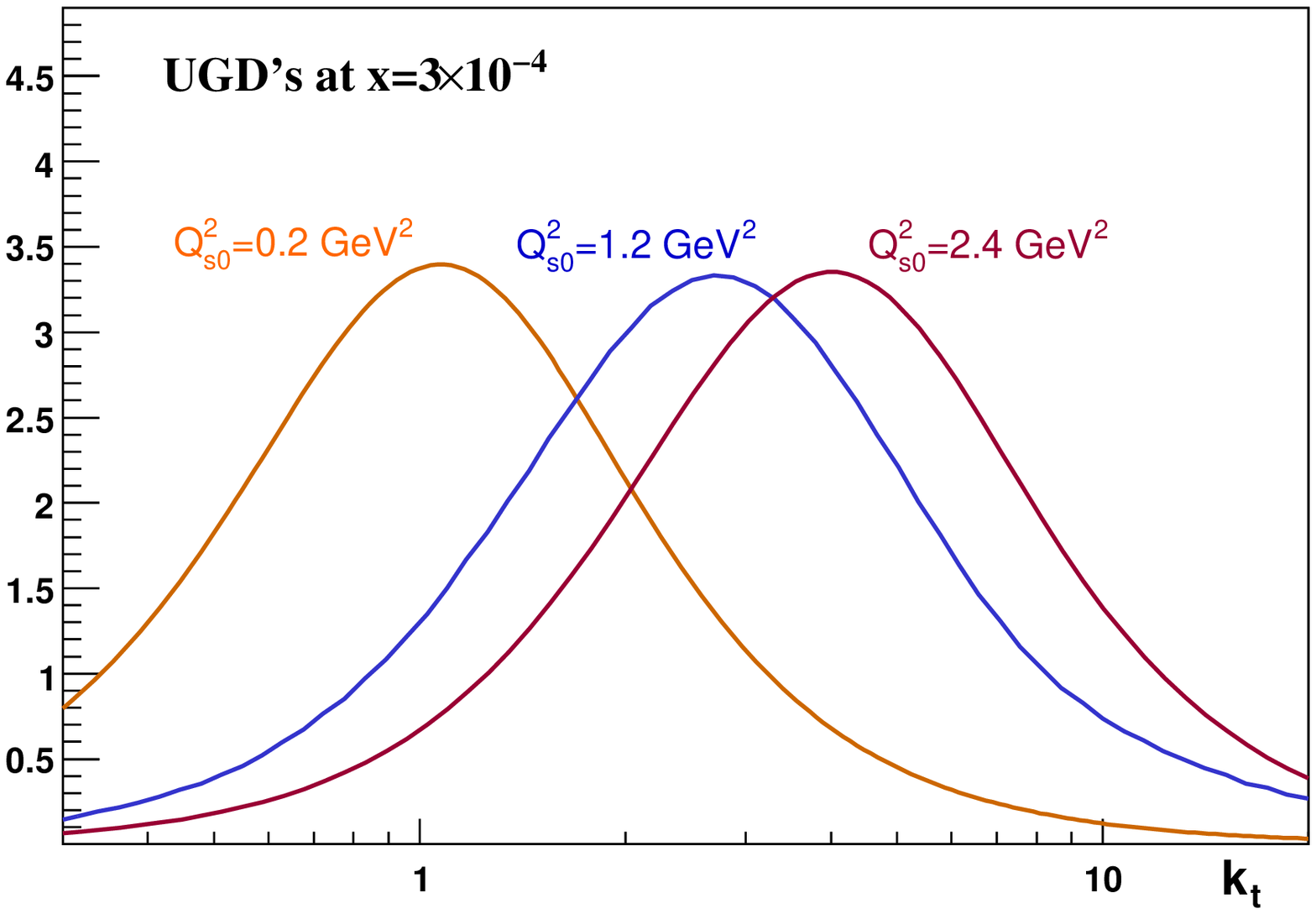}
\includegraphics[width=6cm]{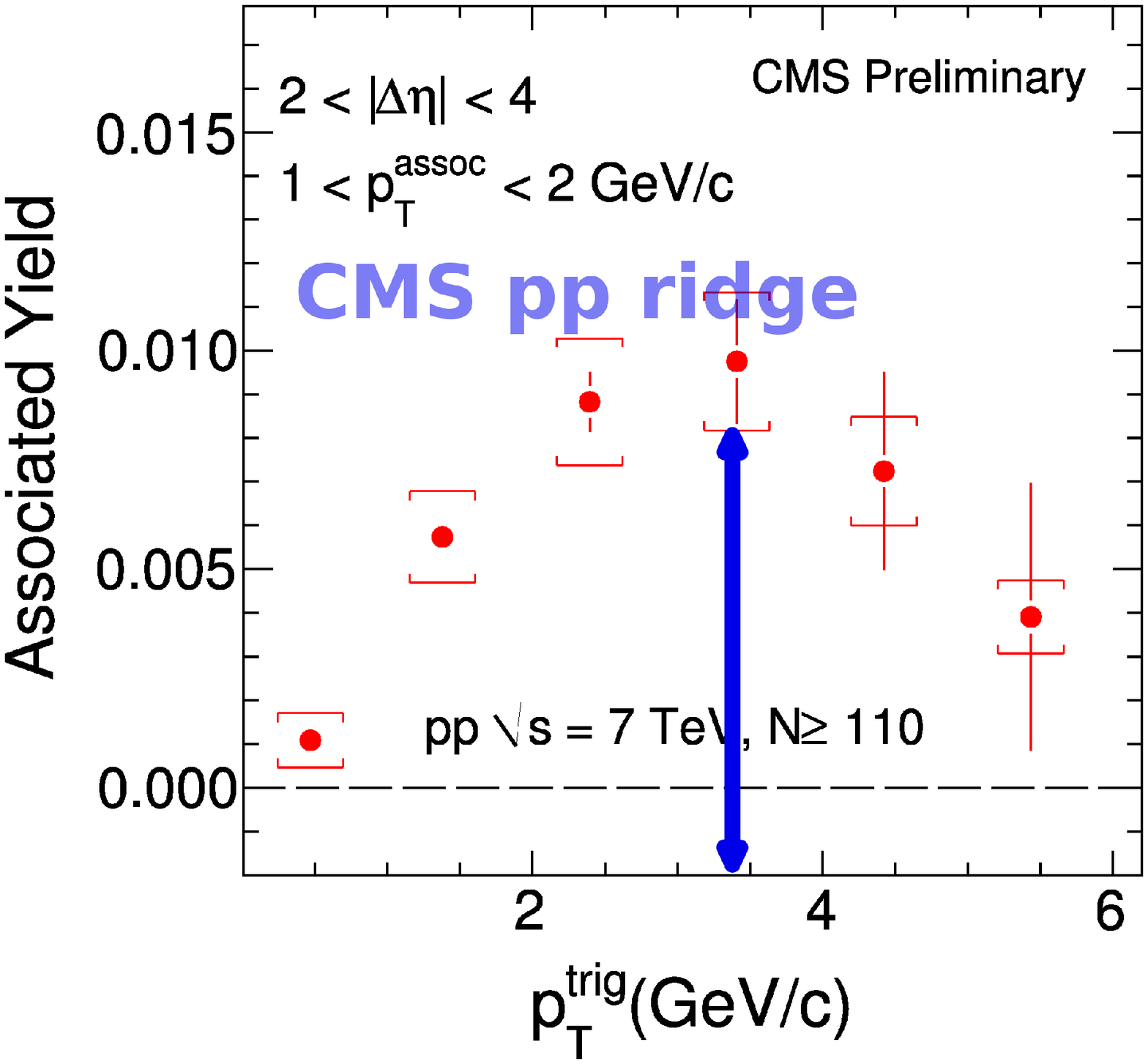}
  }
  \caption{Left: Albacete's unintegrated gluon distribution
    $\Phi(k_\perp,x)$ obtained from rcBK evolution of the MV-model
    dipole, eq.~(\ref{eq:MV}). Right: associated
    yield in CMS ridge peaks about a semi-hard scale.}
  \label{fig:UGD}
\end{figure}
The evolution of observables with energy is described by ``small-$x$''
QCD evolution equations (where $x\sim1/\surd{s}$ is inversely
proportional to the energy of the beams). The two-point function of
two light-like Wilson lines, in particular,
\begin{equation}
  \mathcal{N}(r,x) = 1 - \frac{1}{N_c} \langle {\rm tr}\,
V_r V^\dagger_0\rangle
\label{eq:N}
\end{equation}
determines DIS structure functions and single-inclusive cross sections
in hadronic collisions\footnote{On the other hand, the connected
  contribution to multi-particle production in general involves higher
  $n$-point functions.}. $\mathcal{N}(r,x)$ represents the scattering
amplitude of a color dipole of transverse size $r$.
Its evolution with energy is governed by the
non-linear so-called Balitsky-Kovchegov (BK) equation,
\begin{equation}
  \frac{\partial\mathcal{N}(r,x)}{\partial\log(x_0/x)}=\int d^2 r_1\
  K(r,r_1,r_2) \left[\mathcal{N}(r_1,x)+\mathcal{N}(r_2,x)
-\mathcal{N}(r,x)-\mathcal{N}(r_1,x)\,\mathcal{N}(r_2,x)\right]~.
\label{eq:BK}
\end{equation}
Present numerical solutions (and corresponding applications)
of this evolution equation employ a kernel valid to running coupling
accuracy~\cite{AK},
\begin{equation}
  K(r,r_1,r_2)=\frac{N_c\,\alpha_s(r^2)}{2\pi^2}
  \left[\frac{1}{r_1^2}\left(\frac{\alpha_s(r_1^2)}{\alpha_s(r_2^2)}-1\right)+
    \frac{r^2}{r_1^2\,r_2^2}+\frac{1}{r_2^2} \left(
    \frac{\alpha_s(r_2^2)}{\alpha_s(r_1^2)}-1\right) \right]~.
\label{eqKbal}
\end{equation}
From the Fourier transform of the scattering amplitude for a dipole in
the adjoint representation, $\mathcal{N}_A(r,x) = 2\mathcal{N}(r,x) -
\mathcal{N}^2(r,x)$, one obtains the ``(dipole) unintegrated gluon
distribution'' $\Phi(k_\perp,x)\sim k_\perp^2\;
\mathcal{N}(k_\perp,x)$ which determines the single-inclusive gluon
production cross section. Owing to the non-linear term
in~(\ref{eq:BK}) this function vanishes as $k_\perp\to 0$ and peaks
about a semi-hard scale $\sim Q_s(x)$, as shown in
fig.~\ref{fig:UGD}. Although the direct analogy (so far) is somewhat
murky, it is nevertheless tempting to show it side by side with the
``ridge yield'' obtained by CMS for high-multiplicity p+p collisions
at 7~TeV~\cite{CMS_WeiLiQM11} (also shown by J.-H.~Kim at this
meeting); the associated yield in the ridge evidently peaks about a
semi-hard scale, too, which needs to be understood more clearly (for
some initial ideas to relate the CMS ridge to the properties of
non-linear evolution see refs.~\cite{CGCridge}).

\begin{figure}
  \centerline{
\includegraphics[width=8cm]{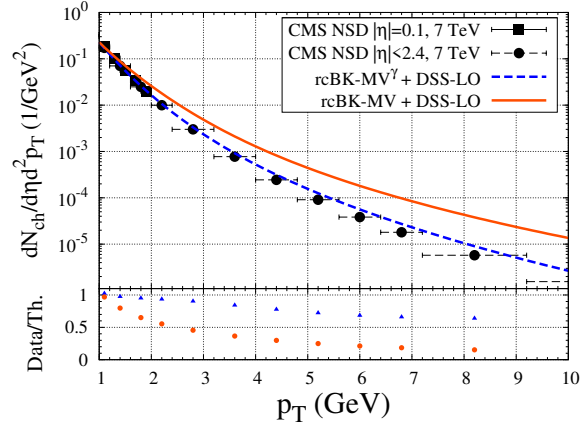}
  }
  \caption{Hadron transverse momentum distributions in the central
    region of $p+p$ collisions at 7~TeV~\cite{DFN}. Theory $K$ factor
    adjusted to fit data at $p_\perp=1$~GeV. CMS data from~\cite{CMSpp}.}
  \label{fig:rcBK_ppLHC}
\end{figure}

As shown in the talks by Albacete, Fujii and Nara, the rcBK
solution also provides a good description of HERA DIS structure
functions at low $x$, of hadron transverse momentum distributions in
the forward region of $p+p$ and $d+Au$ collisions at RHIC, and of
$p_\perp$-integrated hadron multiplicities in $Pb+Pb$ collisions at
the LHC.
\begin{figure}
  \centerline{
\includegraphics[width=7cm]{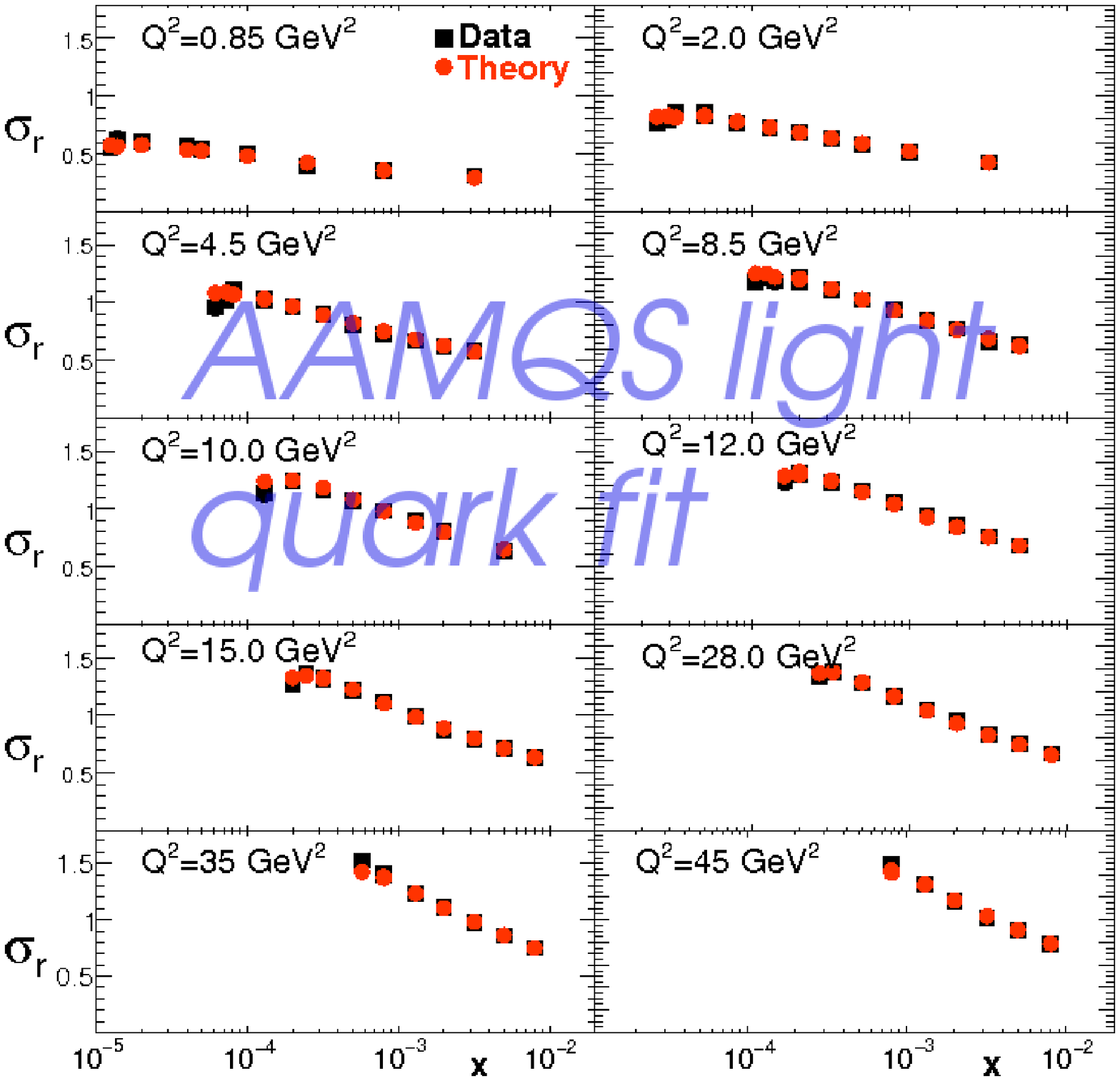}
\includegraphics[width=7cm]{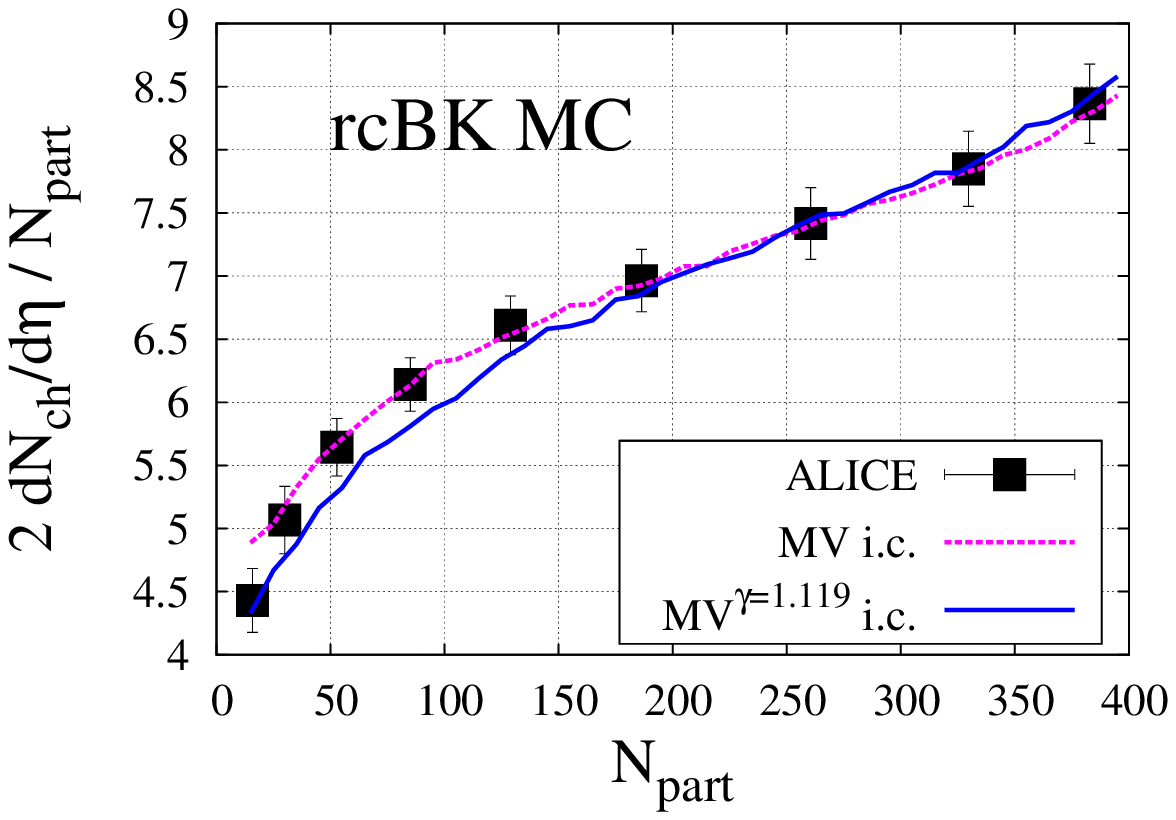}
  }
  \caption{Left: AAMQS rcBK fit to HERA structure functions~\cite{AAMQS}.
    Right: charged hadron multiplicities versus centrality
    for $Pb+Pb$ collisions at 2.76~TeV~\cite{JLAad}. Presented by J.~Albacete.}
  \label{fig:rcBK_F2}
\end{figure}
\begin{figure}
  \centerline{
\includegraphics[width=7cm]{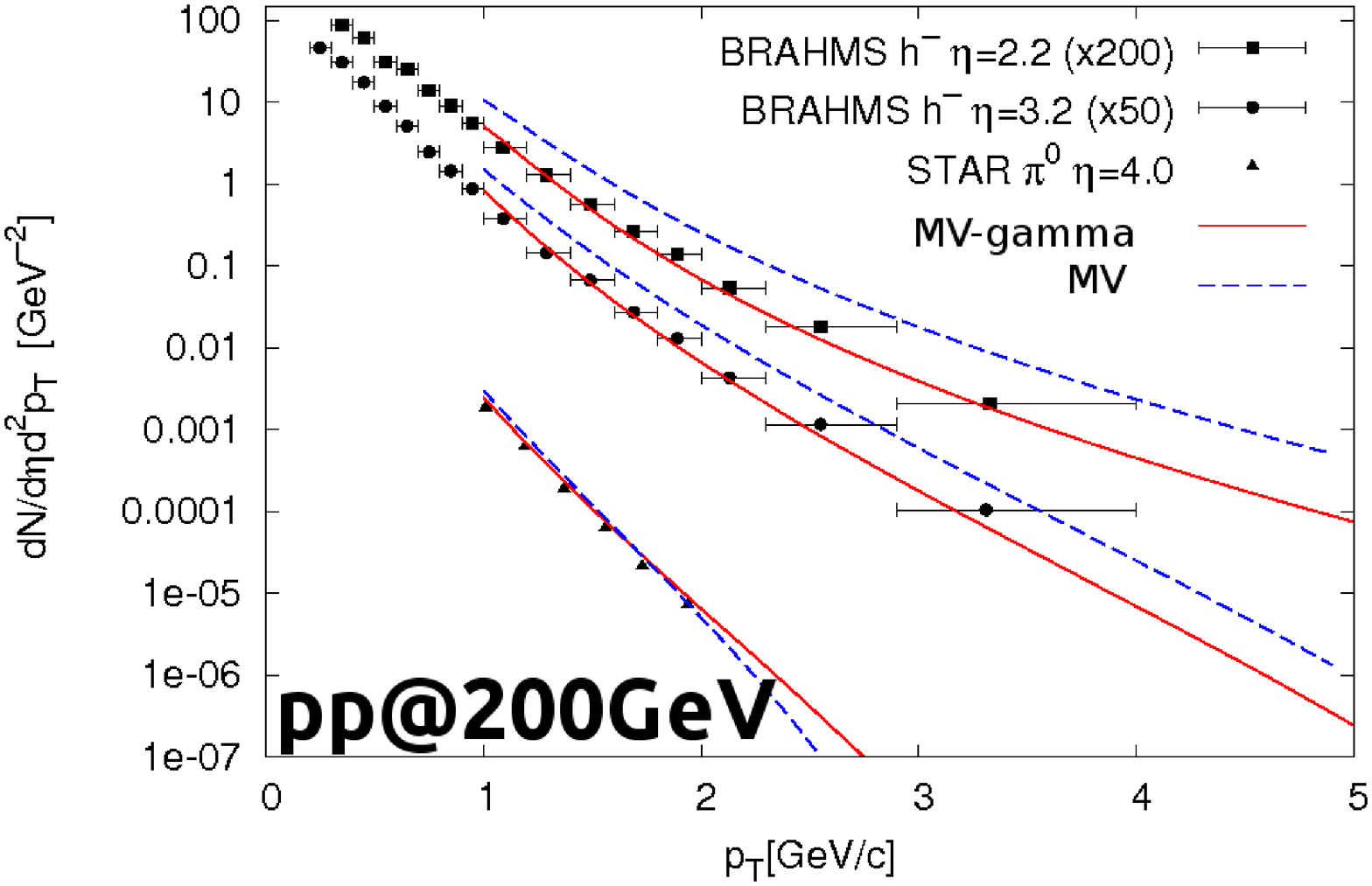}
\includegraphics[width=7cm]{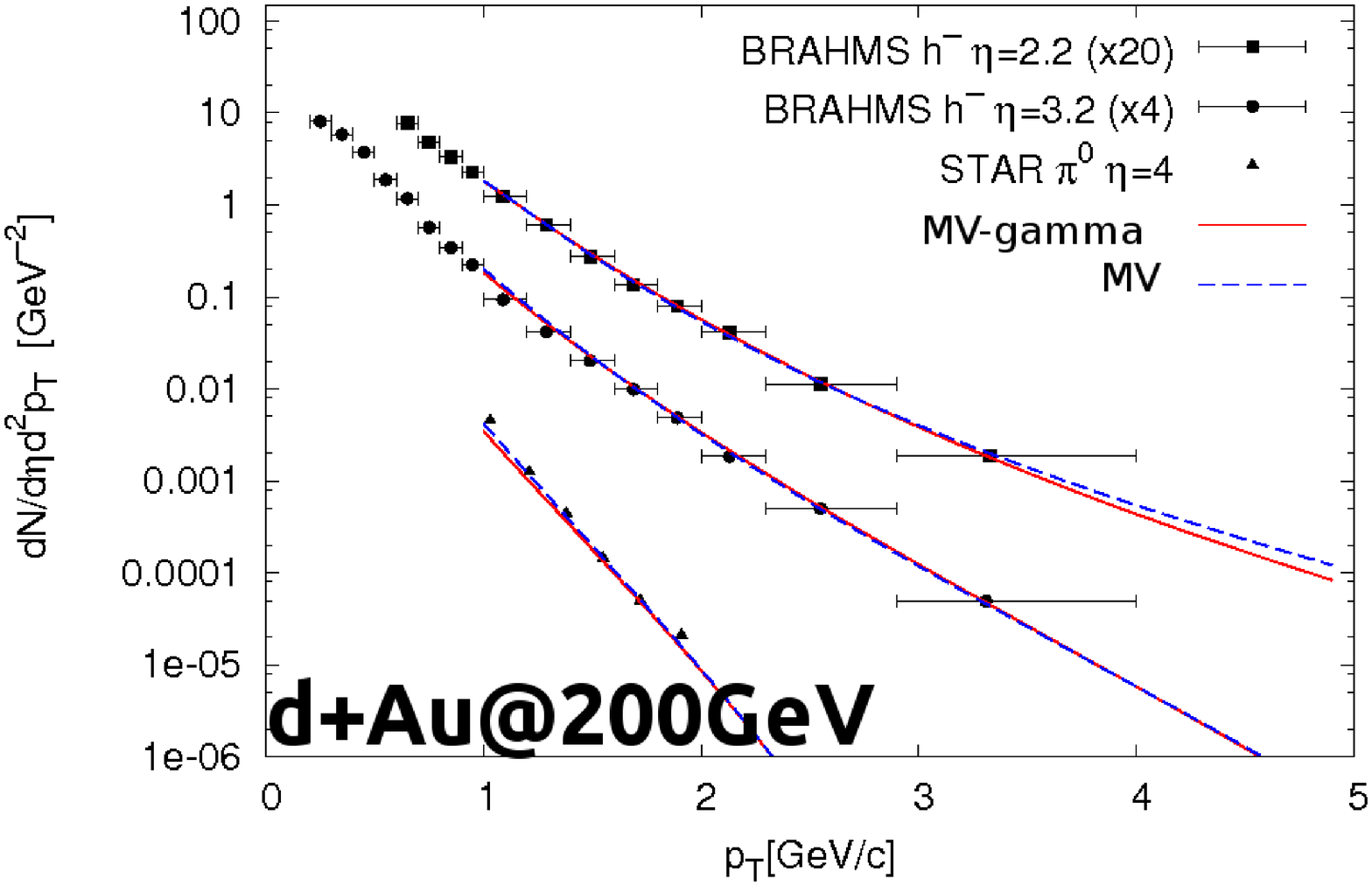}
  }
  \caption{Hadron transverse momentum distributions in the forward
    region of $p+p$ and $d+Au$ collisions at RHIC as presented by
    H.~Fujii at this meeting~\cite{FIN}.}
  \label{fig:rcBK_pp_dA}
\end{figure}
Some examples are shown in figs.~\ref{fig:rcBK_F2}
and~\ref{fig:rcBK_pp_dA}. It is worth noting though that the AAMQS DIS
fit requires an initial condition for eq.~(\ref{eq:BK}) which is not
like the ``MV model'' dipole,
\begin{equation}  \label{eq:MV}
\mathcal{N}_{MV}(r,x_0) = 1 - \exp \left( - \frac{1}{4} \; r^2 Q_s^2(x_0)\;
\log\frac{1}{r\Lambda}
\right)~,
\end{equation}
but one which falls off more rapidly at small $r$:
\begin{equation}  \label{eq:MVg}
\mathcal{N}_{MV^\gamma}(r,x_0) = 1 - \exp \left( - 
\frac{1}{4}\left(r^2 Q_s^2(x_0)\right)^\gamma \log\frac{1}{r\Lambda}
\right)~~~~~~~~,~~~(\gamma >1)~.
\end{equation}
This finding is underlined by the charged hadron transverse momentum
distributions in $p+p$ collisions measured at the LHC, see
fig.~\ref{fig:rcBK_ppLHC}.
The transverse momentum distributions at $p_\perp>Q_s$ have been
obtained by first computing the transverse momentum distribution of
gluons from the $k_\perp$-factorization formula which is then
convoluted with a DSS gluon $\to$ hadron QCD fragmentation
function~\cite{DSS_FF}:
\begin{equation}
\frac{dN_{h^\pm}}{dy\; d^2p_\perp} =
\int \frac{dz}{z^2} \; D_g^{h^\pm}(z,Q^2)\;
\frac{dN_g}{dy\; d^2q_\perp} ~.
\end{equation}
Here, $p_\perp=z q_\perp$ is the transverse momentum of the produced
hadron and the scale $Q^2$ may be chosen as $q_\perp^2$.
For the first time, the extreme energy reach of LHC enables us to
perform quantitative tests of the theory at high transverse momentum
$p_\perp$ but small $x$. Evidently, the MV model initial condition
with rcBK evolution (plus $k_\perp$ factorization) disagrees with the
CMS data. On the other hand, it is comforting to see that the
AAMQS-modified MV$^\gamma$ initial condition which worked for $F_2$ at
HERA provides a good match to the LHC high-$p_\perp$, low-$x$ data as
well.

As is often the case, answers lead to more questions. Such as: is
there any theory which would at least motivate the successful AAMQS
guess~(\ref{eq:MVg})~? The MV model initial condition~(\ref{eq:MV})
can be derived from an effective theory treating the large-$x$ valence
charges as recoilless sources on the light cone (analogous to the
Weizs\"acker-Williams equivalent photon approximation); in the limit
of a very large number of such valence charges, color fluctuations are
described by a quadratic action~\cite{MV}
\begin{equation}  \label{eq:S_MV}
S = \int d^2x_\perp \, \frac{1}{2\mu^2}~\rho^a \rho^a~,
\end{equation}
where $\mu^2(x_\perp) \sim N_{val}$ is proportional to the number of
valence charges at a given transverse coordinate. For a nucleus,
$N_{val} \sim N_c A^{1/3}$ is proportional to the number of colors and
to the longitudinal thickness.

The MV$^\gamma$ initial condition~(\ref{eq:MVg}), on the other hand,
has not been derived (so far) from an effective action for the large-$x$
valence charges. Worse yet, there is no theoretical understanding of
the dependence of the parameter $\gamma$ on $N_{val}$ resp.\ on the
thickness $A^{1/3}$ of a nucleus; $\gamma$ would have to be re-fitted
for every single hadron or nucleus. Understanding the $A$ dependence
of the AAMQS parameter $\gamma$ will also be important for\\
i) the ratio
of $pA$ to $pp$ spectra at the LHC: some estimates using rcBK
unintegrated gluon densities suggest that $R_{pA}$ might be suppressed
at LHC energies, even away from the proton fragmentation region, due
to the effects from small-$x$ evolution~\cite{RpA}; and\\
ii) for predictions or fits of structure functions of heavy ions at
small $x$ which will hopefully be measured by a future electron-ion collider
eIC~\cite{eIC}.

One possibility is that the AAMQS initial condition arises from
corrections to the MV model action which involve higher powers of the
color charge density $\rho$. Such terms are allowed by SU(N)
invariance and involve couplings suppressed by additional powers of
$1/gA^{1/3}$:
\begin{equation}  \label{eq:S_Q}
S = \int d^2x_\perp \left[ \frac{1}{2\mu^2}\rho^a \rho^a
- \frac{1}{\kappa_3} d^{abc} \rho^a \rho^b \rho^c
+ \frac{1}{\kappa_4} \rho^a \rho^a \rho^b \rho^b\right]~.
\end{equation}
Here, $\mu^2 \sim g^2 A^{1/3}$, $\kappa_3\sim g^3 A^{2/3}$, and
$\kappa_3\sim g^4 A$~\cite{DJP}. This action would allow for definite
predictions for the $A$-dependence of the initial condition for rcBK
evolution. It will be interesting to see if~(\ref{eq:S_Q}) is able to
reproduce the essential features of the AAMQS MV$^\gamma$ model.

Coming back to
\begin{equation}  \label{eq:R_pA}
R_{pA}(p_\perp) = \frac{dN_{pA}/d^2p_\perp}{N_{\rm coll} ~ dN_{pp}/d^2p_\perp}~,
\end{equation}
I.~Potashnikova and B.~Kopeliovich in their contribution pointed out
that there are some subtleties in defining the ``number of binary
parton-parton collisions'' $N_{\rm coll}$~\cite{arXiv:1105.1080}. This
is commonly taken as
\begin{equation}  \label{eq:Ncoll}
N_{\rm coll} = \frac{\sigma_{\rm in}^{pp}\, \langle T_A\rangle}
{\langle P_{\rm in}\rangle}~,
\end{equation}
where $T_A$ denotes the thickness function of the target nucleus,
$P_{\rm in}$ is the probability for an inelastic $pA$ interaction, and
$\langle\cdot\rangle$ denotes an average over events. The authors
argued that the cross section in eq.~(\ref{eq:Ncoll}) should be
corrected for the fraction of missed diffractive events. If this
represents a significant fraction then $N_{\rm coll}$ is reduced and
this would affect our interpretation of nuclear effects as measured by
$R_{pA}$.

\begin{figure}
  \centerline{
\includegraphics[width=13cm]{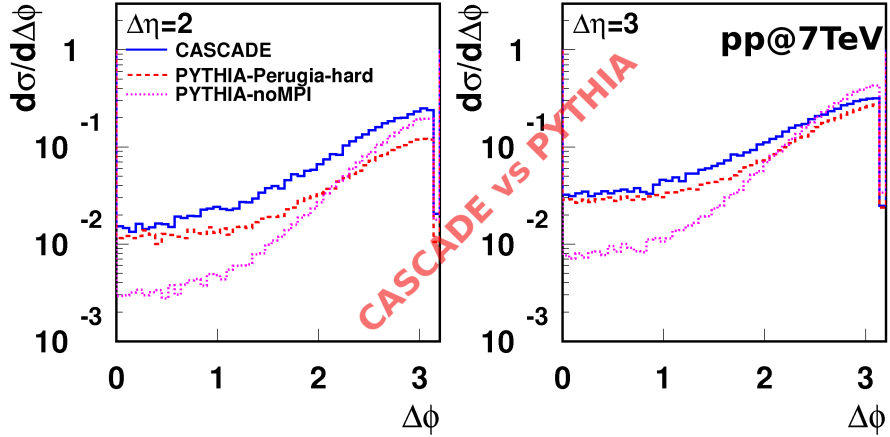}
  }
  \caption{Cross section for two jets with $p_\perp>10$~GeV separated
    by a rapidity interval $\Delta\eta$ and azimuthal angle
    $\Delta\phi$; $p+p$ collisions at $\surd s=7$~TeV. From
    ref.~\cite{arXiv:1012.6037}.}
  \label{fig:JetAzim}
\end{figure}
Another important test for our present knowledge of the unintegrated
gluon densities has been emphasized by K.~Kutak at this meeting. That
is, the single-inclusive cross section at non-central rapidities where
the evolution is asymmetric; also, the angular distribution of
semi-hard jets separated by a rapidity interval
$\Delta\eta$~\cite{arXiv:1012.6037} shown in fig.~\ref{fig:JetAzim}.

\section{Density fluctuations in $p+p$ and $A+A$ collisions}

\begin{figure}
  \centerline{
\includegraphics[width=8cm]{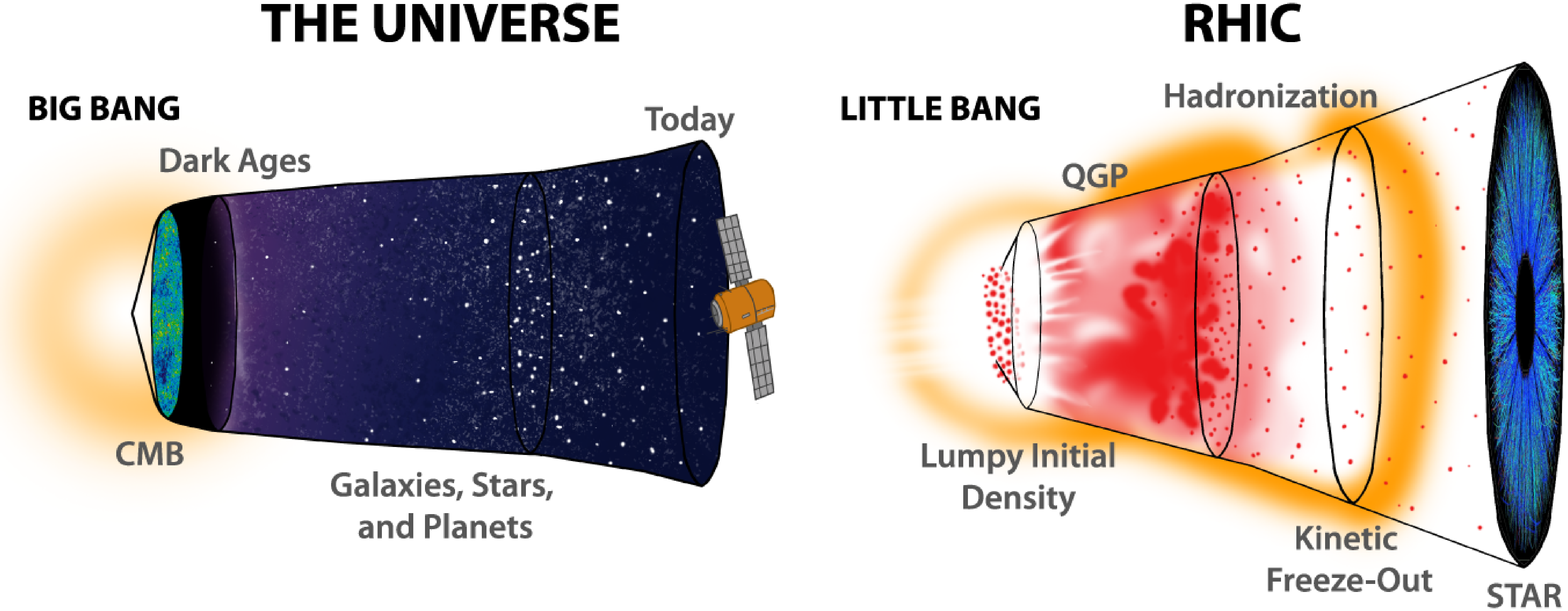}
\includegraphics[width=6cm]{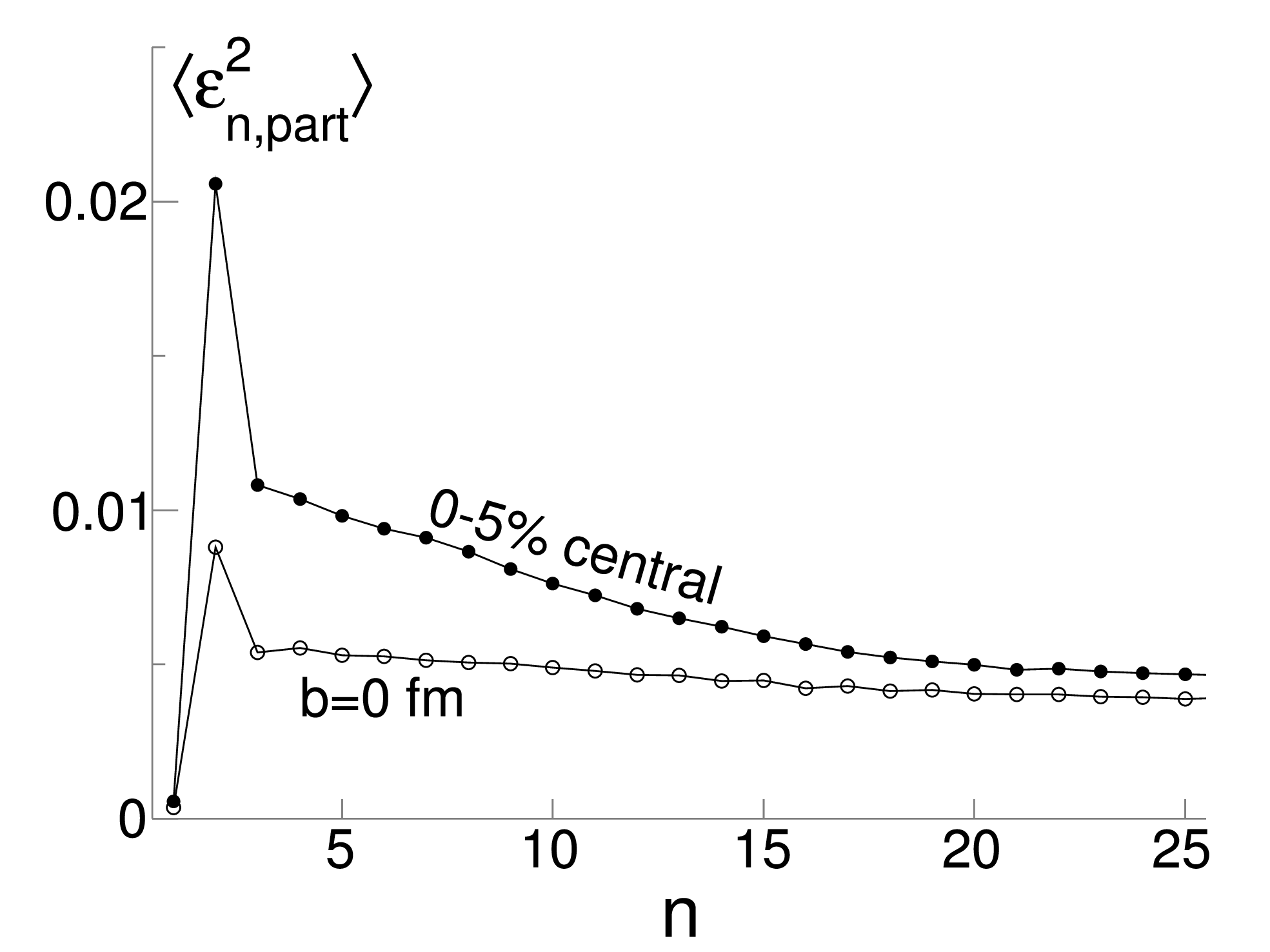}
  }
  \caption{Left: Artists view of the expansion of the universe and of the
    survivial of initial quantum fluctuations as CMB fluctuations
    measured today; and of the expansion of fluctuations in $p+p$ and
    $A+A$ collisions~\cite{MocsySorensen}. Figure by
    A.~Doig. Right: harmonic spectrum of density inhomogeneities in
    central $A+A$ collisions for point-like nucleons.
    Presented at ISMD 2011 by A.~Mocsy and P.~Sorensen.}
  \label{fig:CMB}
\end{figure}

There has been a lot of interest recently in understanding the
magnitude and length scale of fluctuations of the density of produced
gluons in the transverse plane; several presentations at this meeting
addressed related questions. In their talks, A.~Mocsy and P.~Sorensen
noted the analogy to the power spectrum of CMB fluctuations which
provides information on the evolution of the early universe. An
artists view is shown in fig.~\ref{fig:CMB}.
They presented the harmonic spectrum of the initial density distribution
(preceding the hydrodynamic expansion in $A+A$ collisions) in terms of
the eccentricities
\begin{equation}
\epsilon_n^2 = \frac{\langle r^2\cos n\phi\rangle^2 +
\langle r^2\sin n\phi\rangle^2}{\langle r^2\rangle^2} ~.
\end{equation}
$\langle\cdot\rangle$ denotes an average over the initial distribution
of produced gluons in the transverse plane~\cite{BOL}, $dN/d\eta d^2r$. One
commonly considered source of inhomogeneities is due to the
fluctuations of the positions of the nucleons in the nuclei before the
collision. In the absence of a length scale (such as a non-zero
radius of the nucleon) for these fluctuations, and for central
collisions, Sorensen argues that the spectrum of $\epsilon_n$ is
flat\footnote{The peak of $\epsilon_2^2$ in fig.~\ref{fig:CMB} for
  $b=0$ collisions is due
  to the intrinsic deformation of a $Au$ nucleus.}. 

\begin{figure}
  \centerline{
\includegraphics[width=9cm]{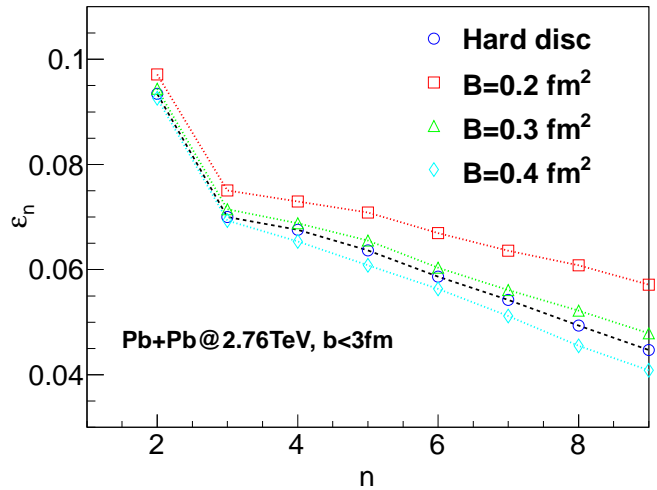}
  }
  \caption{
    Right: Eccentricity spectrum averaged over Monte-Carlo ensembles
    of nucleon configurations with Gaussian valance parton
    distributions, obtained from $k_\perp$ factorization with rcBK
    unintegrated gluon densities. Presented by Y.~Nara~\cite{YN_ISMD}.}
  \label{fig:eps_n_rcBK}
\end{figure}
Y.~Nara showed the Fourier spectrum of eccentricities obtained from a
simulation of heavy-ion collisions based on the $k_\perp$
factorization formula with rcBK unintegrated gluon densities,
averaged over an ensemble of configurations of the colliding
nuclei. Again, here the only source of fluctuations are the random
positions of the nucleons before the collision (the average is
a Woods-Saxon distribution); they determine the locations of the
valence charges for small-$x$ evolution. The valence charge profile
function was assumed to be Gaussian rather than a delta-function. The
Fourier spectrum of eccentricities then drops towards higher
harmonics, the faster the bigger the Gaussian width $B$. A careful
understanding of higher harmonics, in particular of those moments which
are dominated by fluctuations, will require constraints on the
Gaussian width $B$ from exclusive vector meson production in DIS, and from
multiplicity distributions in $p+p$ and $p+A$; see, e.g.,
ref.~\cite{TV} and references therein.

Fluctuations of the nucleon configurations in the colliding nuclei are
very likely not the full story, however, as has been pointed out by
C.~Flensburg and Y.~Hatta. There are intrinsic fluctuations of the
number of emitted gluons even for a fixed number of longitudinally
stacked projectile and target nucleons (or ``participants'') which are
actually amplified in a collision of strong chromo fields~\cite{Glitter}.
Furthermore, one expects fluctuations due to stochastic dipole
splitting events as implemented in Flensburg's DIPSY
Monte-Carlo~\cite{DIPSY}; the probability distribution for splitting
is given by
\begin{equation} \label{eq:Pbfkl}
dP = \alpha_s \frac{N_c}{2\pi^2} \frac{(\vec{x}-\vec{y})^2}
{(\vec{x}-\vec{z})^2\, (\vec{z}-\vec{y})^2}\, d^2z \, dY~,
\end{equation}
if a fixed coupling constant $\alpha_s$ is assumed. Hatta stressed
that~(\ref{eq:Pbfkl}) results in large fluctuations of the
number of dipoles which are strongly correlated in the impact parameter plane.

\begin{figure}
  \centerline{
\includegraphics[width=7cm]{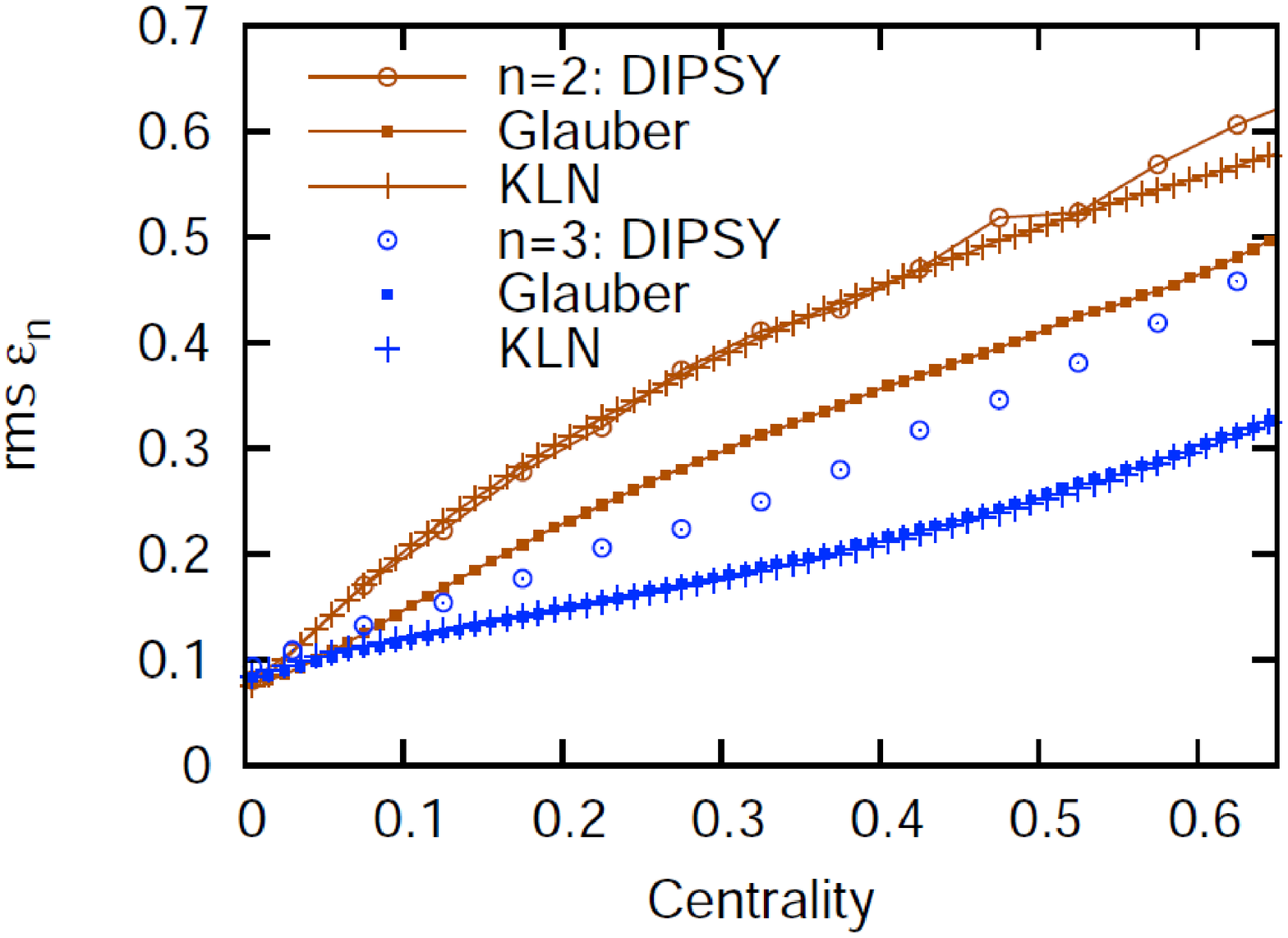}
\includegraphics[width=6cm]{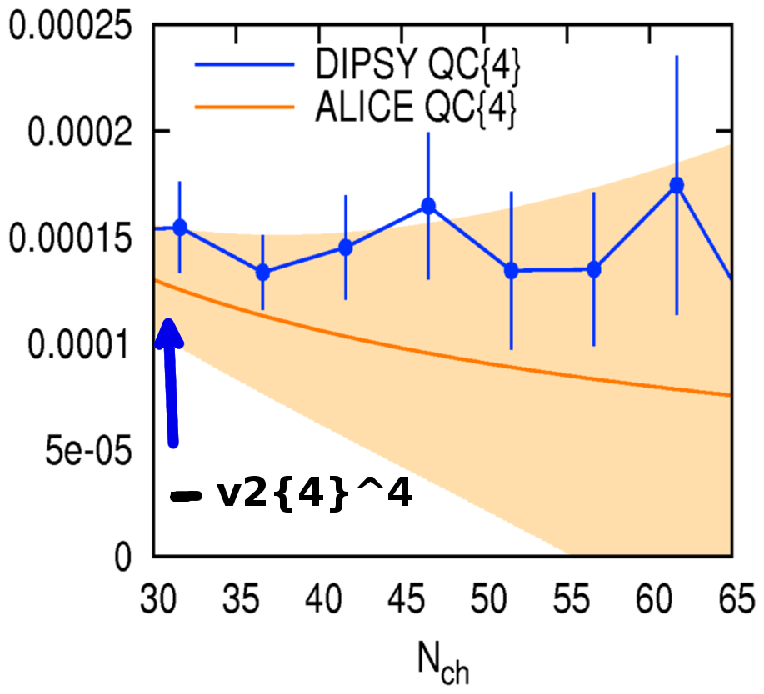}
  }
  \caption{Left: $\epsilon_n$ from the Monte-Carlo stochastic dipole
    splitting model
    DIPSY, from a geometric ``wounded nucleon / Glauber'' model, and
    from Nara's MC-KLN model~\cite{MCKLN}
    (fluctuations of large-$x$ sources only).
    Presented by C.~Flensburg.
  Right: DIPSY prediction for the fourth cumulant of non-flow
  correlations for high-multiplicity $p+p$ collisions at
  7~TeV. Preliminary ALICE data from~\cite{ALICEv24}. Presented by Y.~Hatta.}
  \label{fig:eps_n_DIPSY}
\end{figure}

In contrast, so far solutions of the rcBK equation have been {\em
  averaged} over all random emissions, and corresponding particle
production models have not considered the effects from this type of
fluctuations. Fig.~\ref{fig:eps_n_DIPSY} shows that fluctuations in
dipole splittings can increase $\epsilon_3$ quite significantly, while
$\epsilon_2$ appears to be largely unaffected.

The presence of such large fluctuations in the gluon density is
evident from the multiplicity distribution in $p+p$ collisions at high
energies. It also makes it feasible to define Fourier
``eccentricities'' of the transverse area occupied by produced gluons
in high-multiplicity $p+p$ collisions, in analogy to heavy-ion
collisions. What is measured though are the {\em momentum-space}
azimuthal anisotropies $v_n$ in the final state.  DIPSY predicts that
the large fluctuations would lead to $v_2^4\{4\}$ (measured via
four-particle cumulants) which is {\em negative}, corresponding to
``non-flow'' correlations~\cite{arXiv:1106.4356}.

\section{Summary}

With the ongoing program (and possible upgrades) at RHIC and with the
recent advent of LHC, this is clearly an excellent time for progress
in high-energy QCD. Tools have been constructed, and constantly
refined, to address questions about the strong color fields of hadrons
and nuclei boosted nearly to the light cone and their manifestation in
particle production and in fluctuations analogous to those at the Big
Bang. Hopefully measurements at non-central rapidities as well as for
$p+A$ collisions will be possible at the LHC to probe QCD in a highly
non-linear regime.

\section*{Acknowledgements}

It is a great pleasure to thank the organizers of ISMD 2011 for the
opportunity to visit Miyajima-island (Hiroshima) and to participate in
this interesting multi-faceted conference~!  Also, I gratefully
acknowledge support by the DOE Office of Nuclear Physics through Grant
No.\ DE-FG02-09ER41620 and for PSC-CUNY Award 63382-0042, jointly
funded by The Professional Staff Congress and The City University of
New York.

\end{document}